\documentclass[aps,pra,showpacs,amssymb,nofootinbib,superscriptaddress,twocolumn,longbibliography ]{revtex4-2}
\usepackage{etoolbox} 

\usepackage[T1]{fontenc}
\usepackage[english]{babel}
\usepackage[utf8]{inputenc}

\usepackage{textcomp}
\usepackage{amsmath,amssymb}
\usepackage[colorlinks=true,citecolor=blue]{hyperref}
\usepackage{graphicx}
\usepackage{amsmath}
\usepackage{slashed}
\usepackage[table,xcdraw,dvipsnames]{xcolor}
\usepackage{bbm}
\usepackage{soul}
\usepackage{array,booktabs}
\usepackage{natbib}
\usepackage{amsmath}
\usepackage{bm}
\usepackage{bbold}
\usepackage{hhline}
\usepackage{xcolor}
\usepackage{mathtools} 
\usepackage{float}
\usepackage{tabularx}
\usepackage{tikz}

\usepackage{MnSymbol}
\newcommand{\bmsection}[1]{\par\medskip\noindent \textbf{#1 - }}

\newcommand{\osaFig}[1]{}

\begin{document}

\title{Quantum Information Processing with Spatially Structured Light}

\author{Suraj Goel }
\email{s.goel@hw.ac.uk}
\affiliation{Institute of Photonics and Quantum Sciences, Heriot-Watt University, Edinburgh EH14 4AS, UK}

\author{Bohnishikha Ghosh}
\affiliation{Institute of Photonics and Quantum Sciences, Heriot-Watt University, Edinburgh EH14 4AS, UK}

\author{Mehul Malik}
\email{m.malik@hw.ac.uk}
\affiliation{Institute of Photonics and Quantum Sciences, Heriot-Watt University, Edinburgh EH14 4AS, UK}

\begin{abstract}

Qudits have proven to be a powerful resource for quantum information processing, offering enhanced channel capacities, improved robustness to noise, and highly efficient implementations of quantum algorithms.
The encoding of photonic qudits in transverse-spatial degrees of freedom has emerged as a versatile tool for quantum information processing, allowing access to a vast information capacity within a single photon. 
In this review, we examine recent advances in quantum optical circuits with spatially structured light, focusing particularly on top-down approaches that employ complex mode-mixing transformations in free-space and fibres.
In this context, we highlight circuits based on platforms such as multi-plane light conversion, complex scattering media, multi-mode and multi-core fibers. 
We discuss their applications for the manipulation and measurement of multi-dimensional and multi-mode quantum states.
Furthermore, we discuss how these circuits have been employed to perform multi-party operations and multi-outcome measurements, thereby opening new avenues for scalable photonic quantum information processing.

\end{abstract}

\maketitle

\section{Introduction}

Photons are inherently quantum carriers of information that can travel long distances while interacting minimally with the environment~\cite{o2007optical,krenn2016quantum}. 
For this reason, they are often referred to as “flying qubits”  and are an excellent platform for quantum communication~\cite{divincenzo2000physical,Tissot2024}.
Another major advantage of photonic systems is their low susceptibility to decoherence, allowing single-qubit operations to be implemented with extremely high fidelity. These properties position photons as a key resource for networking quantum computers and realizing distributed quantum information processing~(QIP)~\cite{Slussarenko2019}.

The past two decades have seen significant advances in the field of classical photonic information processing~\cite{wetzstein2020inference, bogaerts2020programmable, shastri2021photonics,  McMahon2023-eu, Bente2025-bt}, largely fuelled by the use of multiplexing across the various degrees of freedom~(DoFs) of light~\cite{xu202111,zhou2022photonic}.
Techniques such as wavelength-division multiplexing~(WDM)~\cite{mollenauer2000demonstration, brackett2002dense, matsushita202041}, time-division multiplexing~(TDM)~\cite{hamilton2002100}, and space-division multiplexing~(SDM)~\cite{richardson2013space, mizuno2017high, puttnam2021space} have been used for decades to achieve massive data bandwidth and throughput for classical communications.
QIP systems also benefit from the same scalability provided by multiplexing,  administered by encoding information in high-dimensional quantum states or \textit{qudits}.

However, there is an important distinction between multiplexing in classical and quantum systems. 
In classical optical systems, the relative phase between different channels is typically irrelevant, as each carries independent information. 
In contrast, qudit multiplexing often involves encoding information across multiple DoFs within the same photon \cite{Malik2016-sr,ciampini2016path,Wang_2018,graffitti2020hyperentanglement}.
Here, maintaining coherence — stable phase relationships between channels — becomes essential in order to preserve quantum superpositions.

Qudits provide several important advantages for quantum networks and quantum computing. From a networking perspective, qudits increase the information capacity per photon and enhance noise robustness in entanglement distribution~\cite{Vertesi:2010bq, mirhosseini_high-dimensional_2015,hu2018beating,ecker2019overcoming, zhu2021high,srivastav2022quick}. In quantum computing, the larger local Hilbert space of qudits enables resource-efficient implementations of gates and algorithms, potentially reducing circuit depth and overhead for fault tolerance~\cite{Ralph_2007,gao2023role,brock2025quantum}. These advantages have motivated extensive theoretical and experimental efforts across a variety of photonic platforms, as well as alternative qudit-capable systems such as trapped ions~\cite{Ringbauer2022-io}, superconducting circuits~\cite{Champion2025}, and atomic ensembles~\cite{ding2016high}.

Realizing the benefits of qudits in practice requires high-dimensional circuits capable of performing arbitrary operations in large Hilbert spaces. 
For polarization qubits, universal linear-optical quantum gates can be constructed using relatively simple building blocks such as beam splitters and wave plates~\cite{o2007optical}. Extension of such techniques for manipulating single qudits use a \textit{bottom-up} approach that involves cascading multiple qubit operations to realize a generalized qudit unitary operation~ \cite{reck_experimental_1994,clements_optimal_2016}. These techniques require the construction of intricate meshes of beam splitters and phase shifters that are usually implemented on integrated photonic platforms~\cite{carolan_universal_2015,wang_integrated_2020,tang_ten-port_2021,taballione_20-mode_2023}. Such approaches present significant technological challenges in scalability that we discuss in the next section. Nevertheless, general operations for path-encoded qudits in dimension up to $d=24$ have been demonstrated recently in integrated photonic platforms~\cite{barzaghi2025low}. 

In general, the control and manipulation of qudits encoded in different photonic degrees-of-freedom (DoF) such as transverse space, time, path, and frequency can be quite difficult, with each DoF bringing its own set of challenges and advantages. Over the past decade, we have seen significant progress in the development of qudit operations across multiple photonic DoFs, including frequency-bins \cite{Lukens:17,Lu:18,Lu:22,Lu:23}, time-bins~\cite{xavier2025energy}, pulse modes \cite{Serino_2023} and hybrid temporal encodings \cite{Serino2025-az, Raymer_2020, Karpinski_2021}. Time/frequency encodings are highly attractive because they offer excellent stability in optical fibers and enable massive parallelization within a single spatial channel. However, their efficient control and manipulation presents significant technological challenges, normally requiring active or nonlinear operations that can suffer from bandwidth and loss issues~\cite{Lukens:17,Lu:23}.

The transverse-spatial photonic degree-of-freedom, herein referred to as spatially structured light, offers a promising alternative for high-dimensional quantum information processing~\cite{Barnett2017-qr,cozzolino2019high,erhard2020advances,forbes2021structured}. The manipulation of spatially structured \textit{classical} light goes all the way back to the development of the first cameras and early telescopes for imaging. Building on this rich heritage, the past two decades have seen the emergence of exciting techniques for controlling spatially structured \textit{quantum} light. For example, unitary and non-unitary transformations for spatially structured quantum light can be realized by using linear optical mode-mixing and passive, phase-only transformations implemented with off-the-shelf devices such as spatial light modulators~(SLMs). 
In this work, we review methods and platforms for building quantum optical circuits for structured light.
We start by discussing the \textit{top-down} design approach that is typically used for constructing optical circuits for structured light, as opposed to the \textit{bottom-up} design commonly used for path-encoded states.
We discuss the various platforms employed to build such circuits by reviewing early work that led to their development. 
Next, we discuss how these circuits are typically used to manipulate and measure quantum light. 
Finally, we review recent work that has employed circuits for spatially structured quantum light in the context of single-party and multi-party operations.

\section{Constructing optical circuits for structured light}

\begin{figure*}[ht!]
    \centering
    \includegraphics[width=\textwidth]{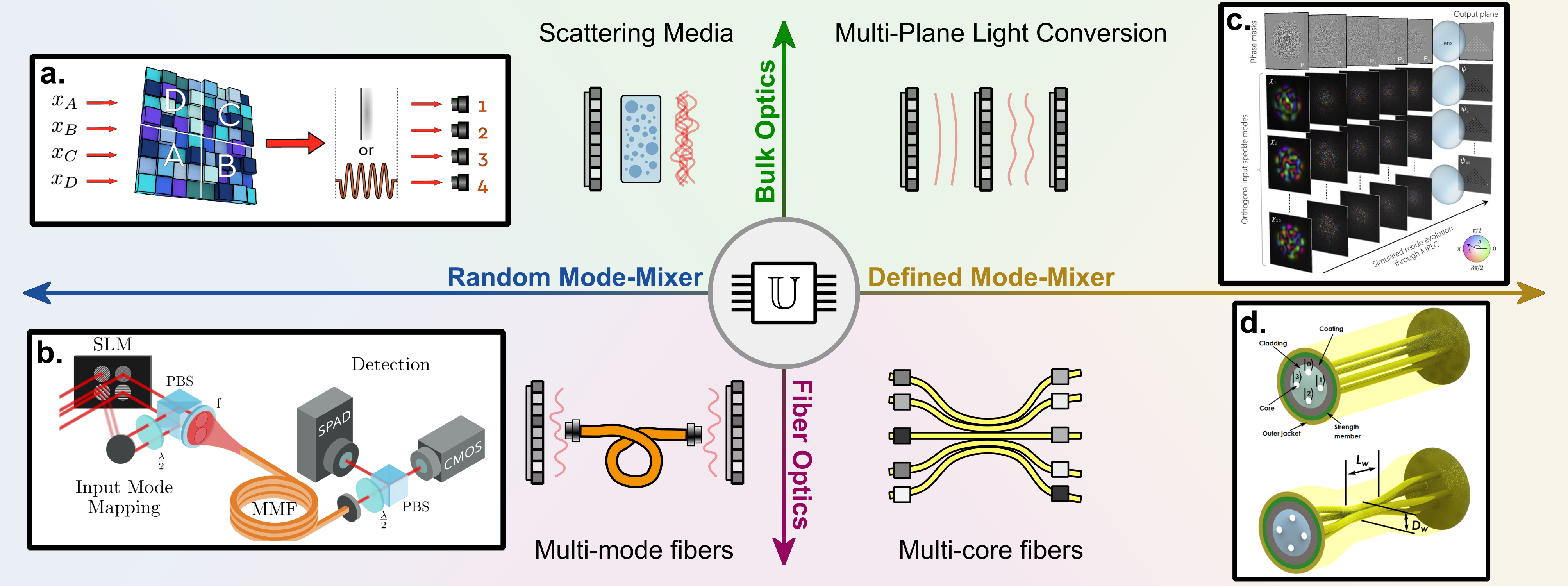}
    \caption{\textbf{Top-down design of reprogrammable circuits based on mode-mixers.} Left (\textbf{a,b}): Circuits based on random mode-mixers such as scattering media (top) and multi-mode fibers (MMFs, bottom) can be combined with programmable phase planes to implement circuits for spatially structured light. 
    \textbf{a.} Random media such as a ground glass diffuser or an MMF is used with wavefront-shaping for optical computing~\cite{matthes2019optical}. Any desired smaller linear operators is extracted from the large random transmission matrix by finding suitable input and output projectors. \textbf{b.} A spatial light modulator (SLM) is used for mapping input modes onto the spatial modes of an MMF. Orthogonal polarizations are combined with a PBS, effectively doubling the number of input channels. After the MMF, outputs can be characterized either by a single-photon (SPAD) detector or a CMOS sensor~\cite{cavailles2022high}. Right (\textbf{c,d}): Devices based on well-defined mode-mixers such as free-space propagation and multi-core fibers (MCFs). \textbf{c.} High-dimensional reconfigurable circuits can be built with multi-plane light converters (MPLCs), which consist of multiple, programmable phase-planes interspersed with free-space propagation~\cite{kupianskyi_high-dimensional_2023}. The inset depicts the simulated behaviour of a 5-plane MPLC for sorting 55 orthogonal speckle modes. The top row shows the phase masks implemented at each plane and the three following rows depict the evolution of three orthogonal speckle modes. 
    \textbf{d.} The inset shows an MCF (top) that can be heated along its length and symmetrically pulled from both ends to create a multi-port beam splitter (bottom) that functions as a mode-mixer for MCF cores~\cite{carine2020multi}. 
    Insets adapted from: \textbf{a,} Ref.~\cite{matthes2019optical} under OSA Open Access Agreement;
    \textbf{b,}  Ref.~\cite{cavailles2022high} under \href{https://opg.optica.org/content/library/portal/item/license_v2}{OSA Open Access Agreement};
    \textbf{c,}  Ref.~\cite{kupianskyi_high-dimensional_2023} under a Creative Commons License \href{https://pubs.aip.org/aip/app/article/8/2/026101/2870744/High-dimensional-spatial-mode-sorting-and-optical}{CC BY};
    \textbf{d,} Ref.~\cite{carine2020multi} under \href{https://opg.optica.org/content/library/portal/item/license_v2}{OSA Open Access Agreement}. }
    \label{fig:fig1}
\end{figure*}

Several methods have been explored to manipulate light in the transverse-spatial degree of freedom. Early advances used bulk optical elements such as dove prisms or specially structured phase plates to implement specific optical transformations for sorting~\cite{leach_interferometric_2004,berkhout_efficient_2010}, transforming~\cite{Malik2016-sr, Krenn2016b,babazadeh2017high}, and measuring \cite{mirhosseini_efficient_2013, mirhosseini_high-dimensional_2015} photons carrying orbital angular momentum (OAM).
In parallel, devices such as q-plates \cite{marrucci2006optical,rubano2019q} were used to couple polarisation with the spatial DoF, allowing access to an even larger range of optical transformations~\cite{d2012complete}. 
Other approaches have harnessed sophisticated modal bases to perform specific transformations and measurements using off-the-shelf optical components such as imaging systems, cylindrical lenses, and optical slits~\cite{Wang2017a,Zhao2019-zd, li2020programmable, solis2021enhanced,koni2024emulating}.

An ideal optical transformation is described by a general unitary transformation $\mathbb{U}$ that maps a set of input optical modes onto a set of output modes~\cite{Miller2012a, Miller2019}. 
The first design of such a transformation for a photonic qudit encoded in distinct optical ``paths'' was proposed by Reck et al.~\cite{reck_experimental_1994}.
They proposed the construction of an arbitrary unitary in a bottom-up manner by using a cascade of Mach-Zehnder interferometers (MZI) and reconfigurable phase shifters to program a generalized unitary transformation. This technique and follow-up work on the bottom-up approach~\cite{clements_optimal_2016, kumar_unitary_2021} has sparked significant progress over the last two decades, enabling the generation, manipulation, and measurement of complex photonic quantum states~\cite{shadbolt2012generating,wang2018multidimensional, wang_integrated_2020,llewellyn2020chip, Maring2024-od, cao2024photonic}. However, significant challenges exist in scaling this approach beyond the 10-mode regime. The number of optical elements required scales quadratically with the dimensionality of the circuit, which introduces challenges in their precise control and calibration~\cite{harris2018linear,bogaerts2020programmable}. 
On-chip implementations of these circuits suffer from high insertion losses (or low efficiency) when coupled to optical fibres or free-space modes ~\cite{taballione_20-mode_2023, quix_datasheet}. 
Additionally, fabrication errors in individual circuit elements propagate very poorly---even small fractional errors in beam-splitting ratios can cause an exponential drop in fidelity of the overall circuit implementation~\cite{miller_perfect_2015,burgwal_using_2017,Pai2018}. 
Another major caveat of this design is that it is usually limited to a planar geometry, which limits scalability to a large number of modes and excludes general transverse-spatial modes. 

In recent years, an alternative, \textit{top-down} architecture for optical circuits has emerged where the circuit is embedded inside an optical system with a larger dimensionality~\cite{morizur_programmable_2010, labroille_efficient_2014, fontaine_laguerre-gaussian_2019, brandt_high-dimensional_2020,carine2020multi, kupianskyi_high-dimensional_2023, makowski2024large, goel_inverse_2024}. 
This is equivalent to adding a large number of auxiliary modes to a given circuit implementation, which can be advantageous in many ways. 
First and foremost, this design allows one to break out of the restrictive planar geometry of the bottom-up design and access the full transverse-spatial photonic degree-of-freedom. While being much more scalable, this also offers compatibility with general transverse-spatial modes. The top-down architecture also separates the control layers from the mixing layers, which eases some of the design requirements for controlling individual circuit elements. Finally, in stark contrast with the bottom-up design, the mixing layer used can be imprecise or even completely random, which reduces or removes the need for strict fabrication tolerances. Such circuits can be inverse-designed within a large, passive mode-mixing optical element such as a multi-mode waveguide~\cite{hashimoto_optical_2005,sakamaki_new_2007}, scattering medium~\cite{huisman2014programming, huisman2015programmable, matthes2019optical}, and even free-space~\cite{morizur_programmable_2010, labroille_efficient_2014, fontaine_laguerre-gaussian_2019,brandt_high-dimensional_2020, kupianskyi_high-dimensional_2023}.
Here, the mode-mixer is placed between two reprogrammable phase planes, which comprises one circuit layer. Cascading several such layers allows one to achieve better circuit performance in terms of fidelity and loss to auxiliary modes (as captured by the efficiency). In general, various optimisation techniques can be employed to program an optical circuit using this approach, such as wavefront matching~\cite{sakamaki_new_2007} and gradient ascent~\cite{kupianskyi_high-dimensional_2023}.
 
Functionally, the top-down design offers a high degree of control over the fidelity and efficiency of the programmed circuit. Numerical studies have shown that optimal performance (in terms of unit fidelity and efficiency) can be achieved when the number of layers is twice the number of circuit modes~\cite{saygin_robust_2020,wang2024ultrahigh,goel_inverse_2024}. However, when the number of layers is less-than-optimal, there is a finite probability that light ends up in the auxiliary modes at the output, therefore leading to loss in the output modes of interest. In this scenario, fidelity can still be increased by increasing the mode-mixer dimensionality. This is due to the addition of several auxiliary modes within the design, which when coupled with the optimisation routine, allow one to fine-tune the balance of fidelity and efficiency without necessarily compromising on each other~\cite{goel_inverse_2024, kupianskyi_high-dimensional_2023}. 

For the top-down design of circuits, the reconfigurability, scalability, and efficiency can largely depend on the type of mode-mixer being used in between different phase layers~\cite{goel_inverse_2024, saygin_robust_2020}.
There are two ways in which the mode-mixers commonly used for top-down circuit design can be classified.
First, one can classify mode-mixers based on their randomness with respect to the spatial basis in which the circuit is implemented. 
This can vary from a completely random mode mixer---such as complex scattering media~\cite{huisman2014programming, huisman2015programmable, Fickler2017, matthes2019optical} and multi-mode fibers~(MMF)~\cite{matthes2019optical,defienne2020arbitrary, leedumrongwatthanakun2020programmable, cavailles2022high, goel_inverse_2024,valencia2025large}---to well-defined mode-mixers such as free-space propagation~\cite{morizur_programmable_2010, labroille_efficient_2014, fontaine_laguerre-gaussian_2019,brandt_high-dimensional_2020, kupianskyi_high-dimensional_2023,wang2024ultrahigh}, multi-mode beam splitters~\cite{saygin_robust_2020, Wang2017a}, and multi-core fibers~(MCF)~\cite{carine2020multi, martinez2023certification, melo2025all}.
Alternatively, mode-mixers can be classified based on the number of auxiliary modes present in these devices. 
On one extreme are bulk-optics mode mixers, such as free-space and bulk scattering media, which can have a potentially unbounded number of modes~\cite{huisman2014programming, huisman2015programmable, Fickler2017, matthes2019optical, labroille_efficient_2014,brandt_high-dimensional_2020, kupianskyi_high-dimensional_2023, wang2024ultrahigh}.
On the other end are fiber-based mode mixers, such as MMFs and MCFs, which have a fixed number of supported modes~\cite{matthes2019optical,defienne2020arbitrary, leedumrongwatthanakun2020programmable, makowski2024large,cavailles2022high, goel_inverse_2024,valencia2025large, carine2020multi, martinez2023certification, melo2025all}. 
Fig.~\ref{fig:fig1} illustrates the two classifications of the top-down design of optical circuits based on mode-mixer properties. 

Numerical studies suggest that a random mode mixer provides a higher degree of reconfigurability as compared to a mode mixer with inherent symmetries when considering top-down circuits with low plane-count, albeit at the cost of efficiency~\cite{goel_inverse_2024}. 
However, this advantage vanishes when a higher plane-count is used.
The community surrounding random complex scattering media has thoroughly explored their potential for programmable optical circuits.

Bulk-scattering media with a potentially unbounded number of scattering modes have been harnessed to implement universal linear optical circuits using a single phase layer~\cite{huisman2014programming, huisman2015programmable, Fickler2017, matthes2019optical}. 
However, more modes are not necessarily always beneficial, as one needs to maintain precise control over each one of them in order to avoid suffering from losses. 
This is where fiber-based approaches come into play, where  MMFs offer a bounded number of modes while retaining properties of random scattering media. 
MMF-based circuits have therefore been central to QIP technologies such as for linear optical quantum processors~\cite{leedumrongwatthanakun2020programmable,cavailles2022high, makowski2024large}, high-dimensional quantum gates and measurements~\cite{goel_inverse_2024}, and quantum networks~\cite{valencia2025large}.

\begin{figure*}[!ht]
    \centering
    \includegraphics[width=\textwidth]{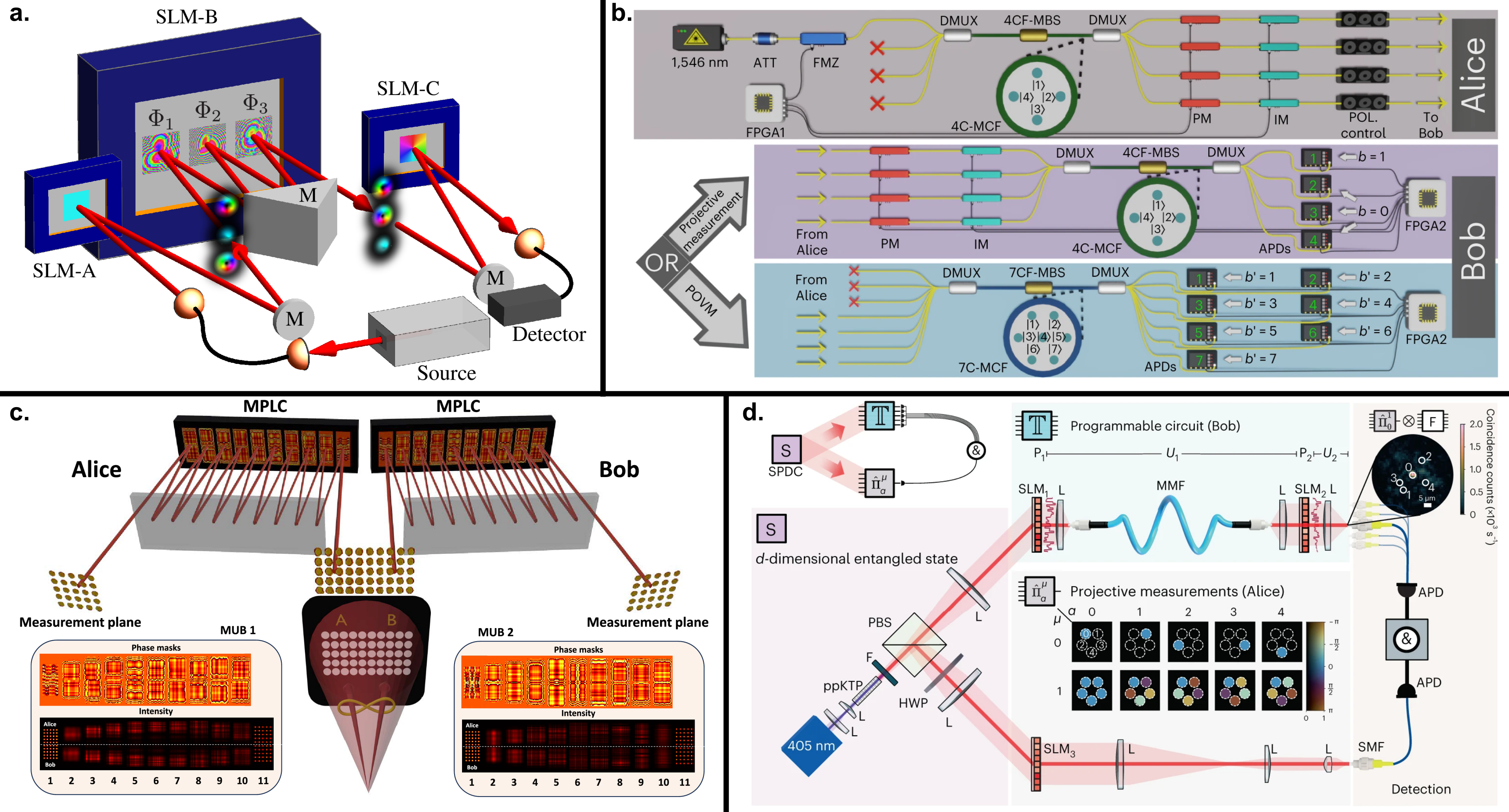}
    \caption{\textbf{Experimental implementations of local quantum operations for structured light.} \textbf{a.} Experimental implementation of a high-dimemsional $\hat{X}$-gate for Laguerre-Gaussian (LG) modes in dimensions $d=3$ with a three-plane multi-plane light converter (MPLC)~\cite{brandt_high-dimensional_2020}. Photons are prepared using a spatial light modulator~(SLM-A), manipulated by the MPLC implemented on SLM-B, and detected via single-outcome projections using SLM-C. 
    \textbf{b.} Implementation of arbitrary POVMs for $d=4$ multi-core fiber (MCF) modes with a four-path interferometer between Alice and Bob~\cite{martinez2023certification}. Alice prepares arbitrary MCF states in $d=4$ with phase modulators and a four-core fibre beam-splitter. 
    Bob performs projective and non-projective measurements using a four-core fiber and a seven-core fiber, respectively.
    \textbf{c.} Demonstration of spatially encoded high-dimensional QKD in a large-scale MPLC~\cite{Lib:25}. Pairs of spatially entangled photons from spontaneous parametric down-conversion~(SPDC) are measured by two parties with a 10-plane MPLC in different mutually unbiased bases (MUBs) of pixel modes. 
    The bottom panels show example transformations of two 25-dimensional MUBs, showing optimized MPLC phase masks and input-mode intensities across all planes. 
    \textbf{d.} An MMF-based optical circuit for manipulating and measuring bipartite quantum entanglement in up to $d=7$ pixel and LG modes~\cite{goel_inverse_2024}. One photon from a pair of spatially entangled photons generated is sent to Alice, who makes projective measurements using SLM$_\text{3}$ and a single-mode fiber (SMF). The other photon goes to Bob, who applies a top-down programmable circuit (an MMF sandwiched between SLM$_\text{1}$ and SLM$_\text{2}$) to perform multi-outcome measurements. The inset shows a coincidence image for a 5-outcome measurement using a Fourier gate. Figure adapted from: \textbf{a,} Ref.~\cite{brandt_high-dimensional_2020}, under a \href{http://creativecommons.org/licenses/by/4.0/}{CC BY 4.0} License; \textbf{b,} Ref.~\cite{martinez2023certification} under a \href{https://arxiv.org/abs/2201.11455}{CC BY 4.0} License; \textbf{c,} Ref.~\cite{Lib:25}, under an \href{https://opg.optica.org/opticaq/fulltext.cfm?uri=opticaq-3-2-182&id=569816}{Optica Open Access Agreement}; \textbf{d,} Ref.~\cite{goel_inverse_2024} under a \href{https://arxiv.org/abs/2204.00578}{CC BY 4.0} License.} 
    \label{fig:fig2}
\end{figure*}

A limitation of complex-media-based optical circuits is that they normally require the transmission matrix of the mode-mixer to be characterized before an optical circuit can be implemented. While there are methods for characterising multiple scattering media separated by phase layers simultaneously~\cite{goel_referenceless_2023,rocha2025self}, they scale poorly for more than a few phase layers. This is where circuit designs harnessing defined mode-mixers come into play. 
Multi-plane light converters (MPLC)~\cite{zhang2023multi}, sometimes also referred to as deep-diffraction neural networks (D2NN)~\cite{sun2023review}, are platforms that follow the top-down design of optical circuits with phase layers separated by free-space mode mixing operations. First discovered over a decade ago~\cite{morizur_programmable_2010,labroille_efficient_2014}, MPLCs were initially developed for space division multiplexing (SDM), where they were used to very efficiently and effectively sort a large number of spatial modes into an array of spots~\cite{carpenter_multi-plane_2017,fontaine_laguerre-gaussian_2019,fontaine2021hermite}. 
The community quickly realized that the use case for MPLCs can also be extended to building optical circuits for spatial modes~\cite{brandt_high-dimensional_2020,hiekkamaki_high-dimensional_2021, kupianskyi_high-dimensional_2023,wang2024ultrahigh, lib2024resource, Lib:25}.
While devices employing MMFs have a record of $d=8$ input modes being simultaneously manipulated with a single phase layer~\cite{cavailles2022high}, MPLCs have been used to perform up to $d=55$ arbitrary mode transformations using $5$ phase layers~\cite{kupianskyi_high-dimensional_2023}. 
Other architectures have utilised optical components such as lenses and half-wave plates between multiple phase-layers to implement programmable circuits~\cite{ammendola2025compactprogrammablelargescaleoptical,derrico2025programmablephotonicquantumwalks}.

More recently, MCFs have emerged as a fiber-based platform with defined mode-mixing for implementing optical circuits~\cite{carine2020multi,carine2021maximizing,martinez2023certification,melo2025all}. 
By virtue of their deterministic fabrication, they provide more controlled mixing albeit with access to relatively few auxiliary modes. This results in devices with lower losses but limited flexibility as discussed above.
MCF-based circuits employ individual phase-shifters for each single mode, instead of a composite phase-layer such as an SLM. 
Such devices have been used for both generation and generalised measurements of photonic states ~\cite{carine2020multi,martinez2023certification}.
A potential advantage of this platform is that it can readily integrate with existing MCF-based prototype quantum networks~\cite{zahidy2024practical,wu2025integration}.

\section{Applications of circuits in QIP with structured light}

Now that we have reviewed the construction of reconfigurable optical circuits for spatially structured light, here we discuss their applications in quantum information processing (QIP). 
Reconfigurable quantum optical circuits are essential for both the manipulation and measurement of quantum photonic states.
Many QIP protocols rely on high-fidelity and high-purity quantum operations and measurements for successful implementation. 
The quality of quantum measurements in photonics often directly depends on the quality of operations, as measurements are typically realized through a generalized transformation followed by photonic detection.
A quantum operation is described by the formalism of quantum processes or channels~\cite{bertlmann2023modern} and can be completely characterized by performing quantum process tomography~(QPT)~\cite{bouchard2019quantum}. However, QPT is quite resource-intensive and scales poorly with the number of modes, necessitating the use of assumptions on the process such as purity \cite{laing2012super} or more efficient characterisation techniques such as witnesses~\cite{engineer_certifying_2024}.

Generalised measurements of quantum states are typically described by the formalism of positive operator-valued measurements~(POVMs)~\cite{nielsen_quantum_2010, barnett_quantum_2009}. The experimental implementation of POVMs for spatially structured quantum states can be quite challenging, and the results of POVMs are usually obtained by sequentially querying multiple copies of a state with several different single-outcome projective measurements~\cite{Bouchard2018-uy,Erhard2018iua}. 
However, this approach is inefficient as it requires at least $d$ independent measurements to reconstruct a $d$-outcome POVM. Thisadditionally opens up the fair-sampling loophole, which is critical for tests of nonlocality~\cite{clauser_proposed_1969, pearle_hidden-variable_1970}.
An optical circuit for spatially structured light allows us to perform arbitrary multi-outcome measurements, thereby implementing a true POVM in a single shot~\cite{goel_simultaneously_2023}.
This is particularly important for quantum communication protocols, where efficiency and unbiased sampling directly impact security.

Photonic circuits can be broadly classified according to their applications for the local or global manipulation and measurement of spatially structured states. An commonly used example of a global measurement is an entangling Bell state measurement, which is a workhorse for many QIP protocols. Although circuit construction methods are largely the same in either case, their implications for QIP vary greatly.

\subsection{Single party operations and local measurements}

Performing \textit{arbitrary} local operations and measurements is the cornerstone of QIP and is essential for communication and computation tasks. 
For instance, in quantum key distribution (QKD), which allows two parties to establish a shared random secret key, local operations performed by each party are indispensable. Both prepare-and-measure and entanglement-based QKD protocols require arbitrary local operations to encode, manipulate, and extract the relevant quantum information~\cite{gisin2007quantum}. Additionally, many QIP protocols rely on local operations and classical communication (LOCC) \cite{Chitambar2014-vz}, which can be useful in transforming entanglement distributed between two or more parties via protocols such as entanglement distillation and dilution~\cite{Bennett1996,Bennett1996_2}, or in protocols for state preparation and discrimination~\cite{chefles1998unambiguous,franke2012unambiguous}.

Early works on the manipulation of spatially structured quantum light employed bulk optical elements such as dove prisms or specially structured phase plates to implement specific single-party high-dimensional quantum gates~\cite{babazadeh2017high}, as well as to perform complementary measurements~\cite{berkhout_efficient_2010,Malik2014ht,mirhosseini_high-dimensional_2015} in the OAM basis. 
The first demonstration of reconfigurable single-party high-dimensional quantum operations for transverse spatial modes employed an MPLC for performing quantum gates in the Laguerre-Gaussian basis~\cite{brandt_high-dimensional_2020}.
The experiment used a preparation and measurement stage to perform QPT and certify high-purity transformations in up to $d=5$ dimensions using a 3-plane MPLC optimised using the wavefront-matching method (Fig.~\ref{fig:fig2}a). 
In parallel, other experiments explored high-dimensional optical circuits with lenses placed between two programmable phase-layers and were characterized using QPT reaching dimensions as high as $d=15$ ~\cite{Wang2017a,Zhao2019-zd}.
While such lens-based transformations are reconfigurable, they are designed for specific spatial mode bases and exhibit intrisic loss~\cite{li2020programmable}.
Optical circuits in dimensions up to $d=55$ have recently been demonstrated for various spatial mode bases with a $5$-plane MPLC optimized using the gradient ascent algorithm~\cite{kupianskyi_high-dimensional_2023}. 
However, this experiment employed off-axis holography to characterise their optical circuits instead of QPT, allowing a quantification of high fidelities under assumptions of purity. 
Another recent experiment employed a $4$-plane MPLC to report fideleties of up to $99.6\%$ by performing QPT for $d=3$ dimensional gates~\cite{wang2024ultrahigh}. 
Moreover, MPLCs were also explored for implementing non-unitary operations and generalized POVMs by employing a $4$-plane MPLC for up to $d=7$ dimensional transformations in the Hermite-Gaussian~(HG) basis~\cite{goel_simultaneously_2023}. 
More recently, a $10$-plane MPLC was used to perform local operations in $d=25$ on bipartite entangled states and using them to demonstrate high-dimensional QKD~\cite{Lib:25} (Fig.~\ref{fig:fig2}c) as well as cluster state quantum computing~\cite{lib2024resource}.
The former demonstration was assisted by the clever geometry of pixel modes, which were used as the choice of modal basis for mutually unbiased bases~(MUBs), which are central to quantum networking protocols.
Essentially, the $25$ pixels were arranged in a $5 \times 5$ grid such that $d=5$ dimensional MUBs could be implemented across each row or column, allowing for a $d=25$ dimensional MUB to be implemented along each axis as shown in the inset of Fig.~\ref{fig:fig2}c.
Similar geometries harnessing subspace operations were employed in the cluster state experiment~\cite{lib2024resource}.
Another recent experiment demonstrated reconfigurable quantum gates for the LG basis 
in up to $d=17$ dimensions using a $7$-plane MPLC~\cite{Dahl2024-kt}. The gates were characterized by measuring coupling matrices in one basis, which entailed an assumption of process purity.

While the MPLC is scalable in principle, higher plane counts are accompanied by higher losses in practice.
As such, complex media have been explored in parallel as an alternative platform to manipulate and measure quantum light~\cite{lib2022quantum}.
Over the last decade, several steps have been taken towards the local control and transport of photonic states through complex media such as multi-mode fibers~\cite{defienne2014nonclassical,valencia_unscrambling_2020,courme2023manipulation}.
Recently, an fiber-based circuit was used for realizing quantum gates for both manipulating and measuring high-dimensional entanglement with only 2 phase-planes~\cite{goel_inverse_2024}.
In this demonstration, reconfigurable transformations were programmed in dimensions up to $d=7$ on OAM as well as pixel bases, and were characterised using QPT.
The key advance in this work was the use of the circuit as a generalized multi-outcome measurement device, and its application in high-dimensional entanglement certification (Fig.~\ref{fig:fig2}d).
Due to the availability of auxiliary modes that allow more general operations, the fiber-based platform has also been used to implement non-unitary operations~\cite{goel_unveiling_2024}.

In addition to circuits based on free-space and complex media, devices based on multi-core fibers (MCFs) have been developed as a generalized measurement and control platform. An MCF-based device was used recently for realizing POVMs for spatial modes encoded in MCF cores in up to $d=7$ dimensions~\cite{martinez2023certification} as shown in Fig.~\ref{fig:fig2}b.
This demonstration certified the non-projective nature of the implemented POVM by violating a semidefinite-programming-based witness~\cite{martinez2023certification}.
Together, these advances demonstrate that the experimental ability to realize arbitrary local transformations—-whether through MPLCs, multi-core fibers, or MMFs—-provides the operational building blocks for LOCC protocols.

\subsection{Multi-party operations and joint measurements}

\begin{figure*}
    \centering
    \includegraphics[width=\textwidth]{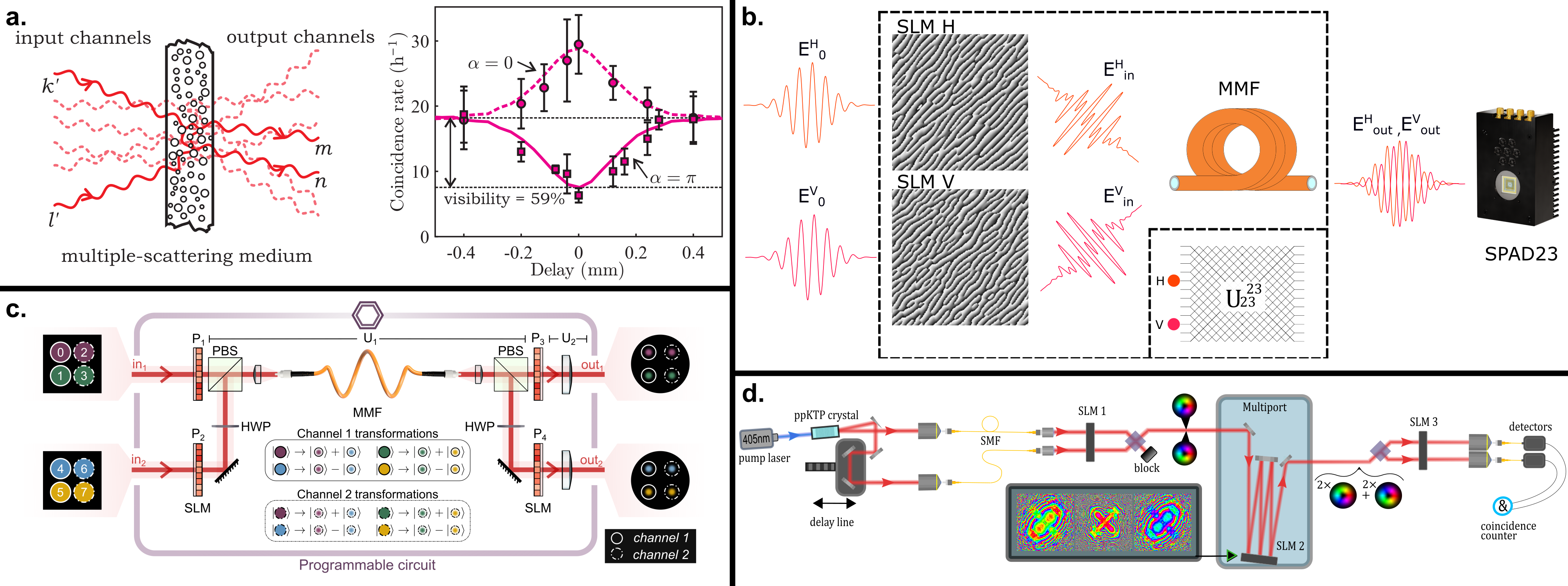}
    \caption{\textbf{Experimental implementations of multi-party operations for structured light}: \textbf{a.} Left: A multiple-scattering medium acts as a multi-mode linear optical network by coupling several input and output channels~\cite{wolterink2016programmable}.
    Right: A plot showing two-photon quantum interference observed by programming a beam splitter into a scattering medium consisting of a layer of white paint. 
    \textbf{b.} A reconfigurable optical network with 2 input and 23 output modes implemented with a spatial light modulator (SLM) and a multi-mode fiber (MMF) used a mode-mixer, with output modes detected using with a SPAD23 camera~\cite{makowski2024large}.
    \textbf{c.} A reconfigurable circuit with 8 input/output modes is constructed with four programmable phase-planes and an MMF~\cite{valencia2025large}. The circuit is used to implement a large-scale, multi-user quantum network with reconfigurable operations for multiplexed entanglement routing, switching, and swapping. \textbf{d.} Two-photon interference is observed between LG modes using a 3-plane MPLC implementing generalized beam-splitters in $d=4$~\cite{hiekkamaki_high-dimensional_2021}.
  Figure adapted from: \textbf{a,} Ref.~\cite{wolterink2016programmable} with permissions from APS;  \textbf{b,} Ref.~\cite{makowski2024large} under an \href{https://opg.optica.org/opticaq/fulltext.cfm?uri=opticaq-3-2-182&id=569816}{Optica Open Access Publishing Agreement}; \textbf{c,} Ref.~\cite{valencia2025large} under a Creative Commons license \href{https://arxiv.org/abs/2501.07272}{CC BY-NC-ND 4.0} with permission; \textbf{d,} Ref.~\cite{hiekkamaki_high-dimensional_2021} with permissions from APS. }
    \label{fig:fig3}
\end{figure*}

While local operations are indispensable for QIP, many quantum information tasks fundamentally require multi-party operations and joint measurements~\cite{zukowski1993event,bennett1993teleporting,bennett1992communication}.
These correspond to global operations acting across multiple subsystems~\cite{pauwels2025classification}, such as the Bell state measurement, which is the archetypical joint measurement.
For instance, linear-optical quantum computing relies on optical circuits to perform entangling gates based on controlled multi-photon interference~\cite{Knill2001-yy}. 
In the context of quantum networks, optical circuits enable the generation of entangled GHZ states which are highly resourceful~\cite{proietti2021experimental-b6c}. 
Similarly, optical circuits are used to link multiple nodes in a quantum network and enable the large-scale distribution of quantum correlations via entanglement swapping~\cite{eisert2000optimal,collins2001nonlocal,kimble2008quantum}. Early work on the development of global operations for spatially structured light focused on circuits constructed with bulk-optical elements \cite{Malik2016-sr}. Interestingly, these efforts went hand-in-hand with the early efforts on computer-designed experiments. The computer algorithm MELVIN was developed to design high-dimensional transformations and experiments for generating complex, multi-photon entangled states with structured light~\cite{Krenn2016b}. One of MELVIN's recipes was subsequently realised for generating the first high-dimensional GHZ-entangled state with photons carrying OAM~\cite{Erhard2018iua}.

Complex media have been studied as a platform of interest for the observation and control of multi-photon interference since the mid-2000s.
This is because complex media support a large number of channels into which photons can naturally scatter, as shown in Fig.~\ref{fig:fig3}a. However, gaining control over this seemingly random process can be quite challenging.
Initial works involved the quantification~\cite{lodahl2005spatial, beenakker2009two} and observation~\cite{smolka2009observation, peeters2010observation} of two-photon interference effects induced by scattering media. Subsequent efforts moved from observing the effects of complex scattering to harnessing it as a controllable resource.
Scattering media such as a layer of white paint and a multi-mode fiber (MMF) were used to implement programmable beam-splitters for light and observe two-photon Hong-Ou Mandel~(HOM) interference~\cite{wolterink2016programmable, defienne2016two}.
Later experiments demonstrated programmable, generalized two-photon circuits in an MMF~\cite{leedumrongwatthanakun2020programmable}, leading to circuits with up to $22$ controllable modes with a single phase layer~\cite{makowski2024large} as shown in Fig.~\ref{fig:fig3}b.
Most recently, an MMF-based reconfigurable optical circuit was used to interconnect two independent quantum networks, allowing operations such as the multiplexed routing and switching of entanglement, as well as multiplexed entanglement swapping between between eight users~\cite{valencia2025large}. 
This was enabled by using 300 spatial-polarization modes in the MMF to embed a programmable 8$\times$8-dimensional linear optical circuit using a two-layer design as shown in Fig.~\ref{fig:fig3}c. The circuit functioned as a programmable quantum interconnect, as well as a multiplexed Bell-state measurement, both of which are fundamental components of a quantum network. 

Scaling MMF-based circuits to multiple layers is experimentally demanding as it involves characterizing multiple sections of MMFs and the use of multiple SLMs. Here, MPLC are a practical alternative as they are able to implement multiple layers using a single SLM. MPLCs have been explored in two recent experiments for implementing programmable multi-photon circuits. 
First, they were used for programming high-dimensional beam-splitter operations in a spatial-mode multi-port for Laguerre-Gaussian (LG) modes in dimensions up to 4~\cite{hiekkamaki_high-dimensional_2021}. The multiport was used for observing generalized two-photon interference effects such as coalescence and anti-coalescence for photons carrying LG modes as shown in Fig.~\ref{fig:fig3}d. Since the device operated entirely in the LG modal space, a third SLM was used to perform spatial-mode projective measurements of the output photons.
More recently, MPLCs were used for implementing spatial-mode beam splitters and observing two-photon interference between single photons emitted from multiple, spatially distinct quantum dots on one cryogenic sample. The MPLC operated on the tilted-plane-wave modal basis, implementing transformations between spatial modes emitted by the quantum dots and measured by a multiple cores of a multi-core fibre~\cite{goel2025controlling}.

\section {Outlook}

With rapid technological advances in QIP, the need for optical circuits operating simultaneously on multiple photons and modes is expected to rise.
Spatially structured light provides a promising pathway to scale these circuits to achieve higher dimensionality, with a range of different possible architectures as discussed above. In recent years, the top-down architecture has emerged as a powerful method for realising such circuits, with implementations in complex scattering media and free-space.
Scalability in the top-down approach can be achieved along two axes: by increasing the number of modes (circuit width) or by increasing the number of phase layers (circuit depth).
However, the efficiency of the circuit drops in both cases, as light is lost due to scattering to auxiliary modes or at phase-layer interfaces.
This challenge particularly affects multi-photon experiments, where the measured coincidences drop exponentially with loss. 
Replacing SLMs with low-loss phase elements such as lithographically etched, fixed MPLCs offers a potential solution at the cost of reconfigurability~\cite{fontaine_laguerre-gaussian_2019, fontaine2021hermite}. However, further work is required to develop architectures that combine the advantages of low loss and reconfigurability.
Additionally, future scalability requires interfacing spatial modes with complementary DoFs of light, i.e. polarization, time, and frequency. 
Progress has already been in the direction -- for example, MPLC-based mode sorters for vector modes that couple spatial and polarization DoFs~\cite{zhang2022polarized,jia2023vector,soma2025complete}, spatio-temporal control using multimode fibers and MPLCs~\cite{mounaix2020time}, and a spatio-spectral mode sorter realized by using an MPLC~\cite{zhang2020simultaneoussortingwavelengthsspatial}.
Nevertheless, further investigation is required to demonstrate arbitrary programmable optical circuits that harness more than two DoFs of light at once.

Finally, scaling the size of optical circuits will necessitate advances in other areas of photonic QIP, such as generation, detection, and storage of spatially structured quantum light. 
While high-dimensional entanglement sources have seen considerable progress recently~\cite{krenn2014generation,valencia2020high,valencia2021entangled}, further developments are needed for realising efficient, multi-photon high-dimensional sources. 
Although single-photon-sensitive detectors with spatial, temporal, and photon-number resolving capabilities for visible wavelengths have seen rapid advances \cite{Roberts2024-dx,hogenbirk2025intensifiedopticalcameratimepix4}, equivalent capabilities at telecom wavelengths remain limited. Aside from a few promising approaches based on single-photon avalanche diode (SPAD) arrays~\cite{brida2009scalable} and recent developments in superconducting nanowire single-photon detector (SNSPDs) arrays~\cite{fleming2025high,stasi2024enhanced}, further advances are required to achieve simultaneous multi-mode detection with photon-number resolution.
Finally, efficient quantum memory schemes capable of storing spatially structured qudits have seen considerable progress in recent years, with the storage of qudits in up to 25 dimensions~\cite{parigi2015storage,Dong2023-bd}. However, significant challenges remain in this area, such as the storage of high-dimensional entanglement, and applications in QIP protocols with feed-forward and synchronization. Nevertheless, the field of quantum information processing with spatially structured quantum light has seen incredible progress over the past 15 years. It is evident that as these techniques rapidly scale, they will deliver significant benefits for quantum technologies across the spectrum.

\bmsection{Funding}
We acknowledge financial support from the European Research Council (ERC) Starting Grant PIQUaNT (950402), the UK Engineering and Physical Sciences Research Council (EPSRC) (EP/Z533208/1, EP/W003252/1), and the Royal Academy of Engineering Chair in Emerging Technologies programme (CiET-2223-112).

\bmsection{Acknowledgments}
We would like to thank Will McCutcheon for helpful discussions.

\bmsection{Disclosures}
The authors declare no conflicts of interest.

\bibliography{references_clean}
\end{document}